\documentclass[trackchanges]{aastex701}

\newcommand{\speed}[1]{#1 km~s${}^{-1}$}

\newcommand{\splitcell}[1]{\begin{tabular}{@{}c@{}}#1\end{tabular}}

\begin{document}

\title{Repeated Sunspot Light Bridge Jets Associated with Slipping Base Brightenings}

\author[0000-0001-9491-699X]{Yadan Duan}
\affil{Yunnan Observatories, Chinese Academy of Sciences, Kunming, 650216, China}
\affiliation{Yunnan Key Laboratory of Solar Physics and Space Science, Kunming, 650216, People's Republic of China}
\email[show]{duanyadan@ynao.ac.cn}
\author[0000-0003-2891-6267]{Xiaoli Yan}
\affil{Yunnan Observatories, Chinese Academy of Sciences, Kunming, 650216, China}
\affiliation{Yunnan Key Laboratory of Solar Physics and Space Science, Kunming, 650216, People's Republic of China}
\email{yanxl@ynao.ac.cn}
\author[0000-0002-6641-8034]{Yuhang Gao}
\affil{School of Earth and Space Sciences, Peking University, Beijing, 100871, People's Republic of China}
\affil{Centre for Mathematical Plasma Astrophysics, Department of Mathematics, KU Leuven, Celestijnenlaan 200B bus 2400, B-3001 Leuven, Belgium}
\email{yh_gao@pku.edu.cn}
\author[0000-0003-4393-9731]{Jincheng Wang}
\affil{Yunnan Observatories, Chinese Academy of Sciences, Kunming, 650216, China}
\affiliation{Yunnan Key Laboratory of Solar Physics and Space Science, Kunming, 650216, People's Republic of China}
\email{wangjincheng@ynao.ac.cn}
\author[0000-0002-9121-9686]{Zhe Xu}
\affil{Yunnan Observatories, Chinese Academy of Sciences, Kunming, 650216, China}
\affiliation{Yunnan Key Laboratory of Solar Physics and Space Science, Kunming, 650216, People's Republic of China}
\email{xuzhe6249@ynao.ac.cn}
\author{Yongyuan Xiang}
\affil{Yunnan Observatories, Chinese Academy of Sciences, Kunming, 650216, China}
\affiliation{Yunnan Key Laboratory of Solar Physics and Space Science, Kunming, 650216, People's Republic of China}
\email{xiangyy@ynao.ac.cn}

\begin{abstract}
Light bridge (LB) jets offer a unique window into small-scale eruptive phenomena within sunspots, the Sun's strongest magnetic environments; however, their generation mechanism remains a subject of debate. Using high-resolution observations from the New Vacuum Solar Telescope (NVST), we investigated six recurrent light bridge jets and the slipping motions of their jet base points (JBPs). Analogous to coronal jets, our observations show that these LB jets are characterized by a preceding JBP followed by a collimated jet spire. The JBP of each repeated jet along the LB displays apparent slipping motion at velocities of 0.6–1.5 $\speed$, which is temporally correlated with quasi-periodic enhanced photospheric horizontal motion of 1.3-6.5 $\speed$. Following the slipping JBPs, the resulting jet spires' fronts display similar slipping behaviors within the upper solar atmosphere. The Chinese H$\alpha$ Solar Explorer (CHASE) reveals Ellerman-bomb-like spectral signatures at the JBPs, confirming that magnetic reconnection is operating at the jet base. Based on these results, we propose that repeated 3D reconnection occurring between the horizontal LB field and the ambient vertical umbral field may drive these LB jets. This process appears to be driven and/or modulated by quasi-periodic horizontal motion fueled by convective upflows and the transport of magnetic flux along the light bridge. This work suggests that some LB jets share a common reconnection-driven mechanism with coronal jets and provides direct evidence of slipping reconnection occurring along the sunspot light bridge.
\end{abstract}


\section{Introduction}
Transient and dynamic jet-like activities ubiquitously populate the solar atmosphere. These intriguing jet-like activities, erupting from the solar surface into the upper atmosphere, are considered potential contributors to coronal heating and solar wind acceleration. These plasma ejections have been observed across various solar regions, including active regions \citep[e.g.,][]{2011Sci...331...55D,2012ApJ...750L..25J,2015ApJ...815...71C,2023ApJ...942L..22D,2025ApJ...985L..12G}, quiet regions \citep[e.g.,][]{2014Sci...346A.315T,2016ApJ...832L...7P,2019Sci...366..890S,2021ApJ...918L..20H,2025ApJ...979..195D}, and coronal holes \citep[e.g.,][]{2014ApJ...796...73H,2015Natur.523..437S,2023Sci...381..867C}. Magnetic reconnection is considered a significant driving mechanism responsible for the energy release in these jet-like activities \citep[e.g.,][]{2007Sci...318.1591S}. In terms of morphology, a standard anemone jet is characterized by a spire and a brightening at its base. These features typically form an inverted-Y or $\lambda$-shaped structure, which is the result of magnetic reconnection between a small-scale bipole and ambient open magnetic fields. Particularly, a category of jetting activity is found to occur on the light bridges of sunspots. These jetting activities ejected from light bridges are called surges, plasma jets, light walls, chromosphere fan-shaped jets, or LB jets \citep[e.g.,][]{1973SoPh...28...95R,2016ApJ...833L..18Y,2017ApJ...835..240S,2017ApJ...848L...9H,2020A&A...642A..44H}. These long-lasting recurring jet-like activities above LBs in sunspots may release or dissipate magnetic energies stored in sunspots to some extent. 

\par
These LB jets can reach a height of 4-38 Mm, and the speeds of these jets are in the range of 50-400 $\speed$ \citep[e.g.,][]{2016A&A...590A..57R,2018ApJ...854...92T}. Using observations from the Interface Region Imaging Spectrograph (IRIS) 1400, 1330 \AA~ slit-jaw images,  a bright front can be observed at the leading edge of these LB jets as they eject plasma upwards \citep[e.g.,][]{2015MNRAS.452L..16B,2015ApJ...804L..27Y,2017ApJ...838....2Z,2017ApJ...848L...9H,2018ApJ...854...92T}. This bright front of jets exhibits oscillation behaviors with a stationary oscillating period of approximately several minutes, which can be interpreted as heating to transition region temperatures by either a shock front or compression. Besides, magnetic reconnection is recognized as another widely driving mechanism of these LB jets \citep[e.g.,][]{2009ApJ...696L..66S}. The magnetic field strength in LBs is approximately 1000 G, which is significantly weaker than the adjacent umbra's strength of about 3500 G \citep{1995A&A...302..543R}. Additionally, the magnetic field lines in LBs are more horizontal compared to those in the umbra \citep[e.g.][]{2014A&A...568A..60L}. These differences create a sharp and strong current layer at the boundary between the horizontal fields of the LBs and the vertical fields of the umbra \citep{2015ApJ...811..137T}. As a result, this boundary may be an ideal site for the occurrence of magnetic reconnection. LB jets are regarded as being produced by magnetic reconnection between a newly emerging flux and a pre-existing magnetic field \citep{2001ApJ...555L..65A}, in the same way as anemone jets \citep{1995Natur.375...42Y}. \citet{2018ApJ...854...92T} reported that many LB jets exhibit the inverted Y-shaped structures, strongly supporting that the classic magnetic reconnection process is the cause of the LB jets. Based on both observations and numerical simulations, \citet{2015ApJ...811..137T,2015ApJ...811..138T} emphasize that the evolution of magnetoconvection plays an important role in driving the brightenings and LB jets. Precisely, the weak horizontal magnetic field is transported from the solar interior to the surface by the convection upflow in the LBs, where it continuously undergoes magnetic reconnection with the strong vertical magnetic field of the surrounding umbra.

Although previous studies have revealed scarce clues of magnetic reconnection driving some LB jets, the three-dimensional characteristics of slipping magnetic reconnection have never been reported before. Here, with joint observations from the New Vacuum Solar Telescope (NVST; \citealt{2014RAA....14..705L,2016NewA...49....8X}), the Chinese H$\alpha$ Solar Explorer (CHASE; \citealt{2022SCPMA..6589602L,2022SCPMA..6589603Q}), and the Solar Dynamics Observatory (SDO; \citealt{2012SoPh..275....3P}), we investigated the physical relationship between these LB jets and the related signals seen in the lower atmosphere, such as the slipping jet bright points at their base and horizontal motion in the photosphere, by capturing their spatiotemporal correspondence. This study (1) confirms that some LB jets share a common reconnection-driven mechanism with coronal jets; (2) provides direct observational evidence of slipping reconnection occurring above LBs; and (3) reveals new clues that photospheric horizontal motion can modulate chromospheric reconnection and trigger 3D slipping reconnection.

\section{Observations} \label{sec:instr}
The data we are interested in was taken on 2025 July 30 above a negative-polarity sunspot in active region (AR) NOAA 14153. NVST high-resolution images at H$\alpha$ passbands cover the complete process of multiple times of slipping motions of the LB jets from 02:41 UT to 04:18 UT, with a temporal resolution of 45 s and a pixel resolution of 0.\arcsec165. The TiO (7058 \AA) passband of the NVST provided information about the horizontal motion on the photosphere LB during the same period, with a temporal resolution of 30 s and a pixel size of 0.\arcsec 052. CHASE tracked AR 14153 from 03:26 UT to 03:55 UT, providing H$\alpha$ imaging spectroscopic observations, with a time cadence of approximately 60 s and a pixel spectral resolution of 0.024 \AA~. The Atmospheric Imaging Assembly (AIA; \citealt{2012SoPh..275...17L}) on board the SDO provides data in seven EUV and two UV channels, with a spatial pixel size of 0.\arcsec 6 and a time cadence of 12 s for EUV passbands and 24 s for UV passbands. The Helioseismic and Magnetic Imager (HMI; \citealt{2012SoPh..275..207S}) on board SDO has a time cadence of 45 s and a pixel resolution of 0.\arcsec 5. In this study, we mainly used 171 \AA~and 1600 \AA~ images from the AIA to study the response of LB jets in the transition region and the lower corona. 
\par
\section{RESULTS} \label{sec:OBS}
\subsection{Overview of Repeated LB Jets}
The HMI longitudinal magnetic field and NVST TiO images show that a light bridge is located above a negative sunspot at S35W40 (Figure 1 (A) and (B)). Figure 1 (C1)$-$(C6) presents six repeated jets above the LB within the observational timeframe covered by the NVST H$\alpha$ and TiO data.  These events are hereafter referred to as case 1 (C1) through case 6 (C6) (see also Table 1). These jets, originating from the LB, exhibit a slipping motion in a particular direction that resembles the fanning of a peacock's tail (see the animation). Consequently, they are categorized as “LB jets'' in this study. A notable characteristic is that significant brightenings occur at the base of the LB jets. Additionally, Figure 1(B) and the corresponding animation illustrate multiple horizontal motions occurring simultaneously beneath the LB jets in the photosphere. To better understand the relationship among the plasma dynamics of the LB jets, the slipping jet bright points in the lower atmosphere, and the horizontal motion in the photosphere, we perform an in-depth analysis of three representative LB jet events (C1, C3, and C4) selected from six recurring LB jets observed at the same location.

\subsection{Lower Atmospheric Signatures and Plasma Dynamics of Recurrent LB Jets}
\subsubsection{The First LB Jet (C1)}
Figure 2 displays a close-up view of the LB jet of one selected example, C1, in different solar atmospheric layers and the Doppler proxy maps. The Doppler proxy maps can be constructed by D=B-R/B+R, where D represents the Doppler proxy maps, and B, R are the emission intensities of H$\alpha$ blue (red) wing \citep[e.g.,][]{2008ApJ...673.1201L,2020ApJ...902....8C,2022A&A...659A..76W,2025ApJ...981..139Y}. This case started with a bright point above the LB at 02:43:17 UT in H$\alpha$ blue wing. At 02:44:45 UT, another bright point appears at the western boundary of LB. It's interesting to note that at 02:46 UT, the pre-existing bright point shows an inverted-Y shape, and the western bright point exhibits discernible slipping motion along the LB. Simultaneously, the beam-like ejecting cold plasma along the collimated magnetic field lines forms the LB jet spires above the JBPs. This inverted-Y morphology and the observed evolutionary pattern, where the JBP appears as a precursor of the jet ejection, are consistent with the launch of coronal jets \citep[e.g.,][]{2016SSRv..201....1R,2021RSPSA.47700217S} and macrospicule \citep[e.g.,][]{2023ApJ...942L..22D}. The JBP at the base of the LB jet is most likely the signature of magnetic reconnection, as suggested by previous observational studies \citep[e.g.,][]{2006A&A...453.1079J,2015A&A...584A...1L,2018ApJ...854...92T}. Additionally, the triggering mechanism is analogous to that of coronal jets, being triggered by magnetic reconnection between the emerging flux and the surrounding open field. These results indicate that the slipping motion of the observed JBPs serves as a suggestive signature of slipping magnetic reconnection occurring at the sunspot LB. 
\par
By 02:46:13 UT, the slipping JBP moves towards the pre-existing JBP, and at 02:46:57, they merge. The merging process of two JBPs can also be observed in AIA 1600 \AA~images in panel (E), with a time delay of roughly two minutes relative to the corresponding signatures captured in the H$\alpha$ wing. Following the merging of the JBP and the slipping JBP, the emission intensity of the latter attains its peak at 02:50:38 UT in the 1600 \AA~channel. In panel (C), the length of the ejected LB jet is notably enhanced following the merger of the two JBPs at 02:48:24 UT. At 02:53:30 UT, the upward-ejected LB jet discontinues along with the dissipation of the associated JBP. This observation feature supports the idea that the LB jet was caused by a reconnection, where the ejection of the LB jet loses its impetus due to the waning or cessation of the magnetic reconnection. Particularly, Figures 2 (A) and (B) show a simultaneous horizontal motion on the LB beneath the slipping JBP.  It is evident that the horizontal motion coincides with the slipping direction of the JBP in a specific direction. These horizontal motion, originating from the northern LB and migrating southeastward, appear to determine the slipping trajectories of the JBP. 
\par
In the Doppler proxy map, the ejected LB jet displays a blueshift signal, indicating an upward motion of the chromospheric plasma. As seen by the black arrows in Figure 2 (D), the JBP and the slipping JBP show redshifted signatures. This redshifted signature, which may be caused by downward flows, has previously been reported at the base of a polar jet \citep{2007PASJ...59S.757K} and an active region jet \citep{2008A&A...481L..57C}. At 02:53:30 UT, this clear redshifted signature from the LB jet is probably the result of the chromospheric plasma falling back to the solar surface along the magnetic field lines, which is consistent with previous observations \citep[e.g.,][]{2024ApJ...962L..38D}. 

 \par
Signatures of the LB jet can also be identified from simultaneous observations of AIA 171 \AA~images, but the JBP and the slipping JBP remain undetected in this passband. In contrast, a slipping motion of brightened plasma is evident at the leading edge of the LB jet spire (indicated by the red circles in panel (F)), first appearing at 02:47:21 UT. By tracking this feature, we observed that it moves along the leading edge of the LB jet in a southeasterly direction.

\subsubsection{The Third LB Jet (C3)}
Figure 3 shows the LB jet (C3) from 03:08 UT to 03:19 UT in NVST H$\alpha$ -0.6 \AA~, Doppler proxy maps, and AIA 171 \AA~images, along with the JBP in 1600 \AA~and the photospheric apparent motion in NVST TiO. The ejection of the LB jet coincides with the evolution of the slipping JBP, beginning upon its appearance and stopping with its disappearance (Figure 3 (C)). This observational result aligns with the findings in C1, reinforcing the idea that the dynamics of the LB jet are driven by magnetic reconnection. A southeasterly slipping JBP along the LB is also discernible in the 1600 \AA~passband shown in Figure 3(E). The LB jet initially exhibits a blueshift signal, which is subsequently followed by the appearance of a redshift signal; notably, the slipping JBP also displays the redshift signature, as shown in Figure 3(D). The slipping motion of the brightened plasma along the leading edge of the LB jet spire persisted for approximately 7 minutes in the AIA 171 \AA~ channel, as denoted by the red circle shown in Figure 3 (F). An associated horizontal motion in the photosphere, co-directional with the slipping JBP, is traced by the cyan circle shown in Figures 3 (A) and (B). The slipping trajectories of the JBP appear to be determined by southeastward horizontal motion. 

\subsubsection{The Fourth LB Jet (C4): Spectroscopic Observations}
A supplementary observation of the LB jet is obtained by CHASE during the period of 03:26–03:55 UT (C4). To investigate the spectroscopic characteristics of the slipping JBP, we examined the CHASE H$\alpha$ data specifically within the LB region, as outlined by the white dashed lines in Figures 4(a)–(f). Figure 4 (c) shows the Doppler map reconstructed from the CHASE data using the weight-reverse-intensity method \citep[e.g.,][]{2016ApJ...816...30S,2018ApJ...863..180W,2024ApJ...962L..38D,2025ApJ...983..143C}. The LB jet exhibits a blueshifted signature, which is consistent with the observations presented in the above NVST Doppler proxy map. Figure 4 (g)$-$(f) shows H$\alpha$ profiles at three different times corresponding to the appearance (panel (d)), peak(panel (e)), and decay (panel (f)) of the JBP. As illustrated in Figures 4(g) and (h), the profiles location of the JBP are characterized by a pronounced intensity enhancement in the H$\alpha$ line wings ($\sim$ -0.8 \AA), along with a weak enhancement in the H$\alpha$ line core, similar to the profiles of Ellerman bombs (EBs)\citep{1917ApJ....46..298E,2016ApJ...824...96T,2017ApJ...838..101H,2019ApJ...875L..30C}. EBs are regarded as the signature of magnetic reconnection between emerging magnetic flux and the background field in the lower atmosphere \citep[e.g.,][]{2007A&A...473..279P,2024A&A...685A...2C}, typically occurring in the upper photosphere or lower chromosphere \citep{2001ChJAA...1..176C}. 
This spectral signature reinforces the findings from C1 and C3, strengthening the interpretation of LB jets being driven by magnetic connection.

\subsection{Dynamic Characteristics and Parameters of LB Jets, Horizontal Motions, and Slipping JBPs}
To investigate the dynamical response of the different solar atmospheric layers during the LB jets evolution, we constructed the time-distance diagrams along Slices C1$-$C4 and S1$-$S4 (see Figures 2 and 3). Figures 5 (A1) and (B1) show the photosphere horizontal motion along the LB at different times for C1 and C3, respectively. The corresponding evolution of the slipping JBPs, as captured in H$\alpha$ off-band and 1600 \AA~channels, is presented in Figure 5 (A2) $-$ (B2), (A3)$-$(B3). A temporal lag is observed between the two channels: the slipping JBPs in the AIA 1600 \AA~ images appear approximately 2 minutes later than their H$\alpha$ off-band counterparts.  Furthermore, in C1, the merger of two JBPs occurs at 02:49 UT in the 1600 \AA~ channel, which similarly lags the H$\alpha$ off-band observation (02:47 UT) by 2 minutes. The 2-minute delay observed between the H$\alpha$ off-band and AIA 1600 \AA~emission potentially reflects the characteristic thermal response time or the interval required for plasma heating to propagate upward through the solar atmosphere. Such an evolution indicates that the magnetic reconnection height increases progressively, likely driven by the continuous emergence of magnetic flux at the JBPs, a scenario consistent with the evolution of bright footpoints in macrospicules reported by \citet{2023ApJ...942L..22D}. Time-distance diagrams along slices C4 and S4, as shown in Figures 5(A4) and (B4), were derived to track the evolutionary dynamics of the LB jet's leading edge. The observations reveal highly dynamic plasma evolution, where emission-enhanced features undergo slipping motion along the LB jet spires' leading edge. 

\par
Based on a detailed analysis of six LB jets, we quantified the horizontal motion velocities, the slipping speeds of the JBPs, and the slipping speeds of the plasma at the jet spires' leading edge, with the results summarized in Table 1. The photospheric horizontal motion velocity ranges from 1.3 to 6.5$\speed$, with a mean value of 3.55$\speed$. This horizontal motion velocity exhibits a progressive increase, peaking at 6.5 $\speed$ around 03:08 UT before subsequently declining.  The slipping speeds of the JBPs vary between 0.6 and 1.5 $\speed$(average 1.1 $\speed$), with a mean lifetime of 12 minutes. Regarding the plasma moving at the leading edge of the LB jet spires, the first event reached a maximum speed of 3.7 $\speed$—significantly exceeding the velocities observed in subsequent events ($\sim$1.1 $\speed$)—and showed a continuous deceleration over time. In alignment with previous studies \citep[e.g.,][]{1973SoPh...32..139R, 2001ApJ...555L..65A, 2009ApJ...696L..66S}, we also determined the ejection speeds and maximum lengths of the LB jets. The average ejection speed was measured at 16.3 $\speed$, with a mean maximum extension of 8.2 Mm from the JBP to the leading edge of LB jet spires.

\par
The spatial distribution of the slipping JBPs is clearly resolved in Figure 6(c) and Animation 1, which reveals that these features are predominantly situated along the boundaries of the LB. This spatial preference prompted a detailed analysis of the magnetic configuration of the entire event. To this end, we utilized the vector magnetogram from the SDO/HMI SHARPs product (hmi.sharp\_cea\_720s)\citep{1994SoPh..155..235M,2009SoPh..260...83L,2011SoPh..273..267B,2014SoPh..289.3549B}, where the tangential components are illustrated in Figures 6 (b). The transverse magnetic field (green arrows) is superimposed on the longitudinal magnetic field map. It can be observed that the transverse field is relatively vertical within the sunspot, undergoes a significant orientation change across the LB, as indicated by the red circle in Figure 6 (b). This configuration implies the presence of magnetic shear between the horizontal fields of the LB and the near-vertical fields of the ambient sunspot, consistent with previous studies \citep[e.g.,][]{2009ApJ...696L..66S, 2015ApJ...811..137T, 2019ApJ...870...90B, 2023ApJ...952...40J}. Additionally, the photospheric flow field was derived from the NVST TiO images using a dense optical flow method \citep[e.g.,][]{2025ApJ...986L..15X}. This approach, implemented via the OpenCV library (cv2. calcOpticalFlowFarneback) based on Gunnar Farneback’s algorithm \citep{10.1007/3-540-45103-X_50}, employed a window size of 11 pixels, corresponding to a spatial scale of approximately $0.\arcsec6$. Observations reveal that the motion originates near the northwestern boundary of the LB with an upward trajectory, as indicated by the cyan arrows in Figure 6(c). Subsequently, the motion redirects southeastward along the LB axis, thereby constituting the pronounced horizontal motion observed in the photosphere. 

\par
To rigorously establish the physical connection between the JBPs and the underlying photospheric horizontal motion throughout the observation period (02:41–03:55 UT), we constructed time-distance diagrams along slices AB and CD (see Figures 6(a) and (c)). The kinematic analysis (Figures 6(e) and (f)) reveals a remarkable one-to-one correspondence and synchronized $\sim$10-minute quasi-periodicities between the six horizontal motion events and the slipping JBPs. This tight coupling is explicitly illustrated by the yellow dashed arrows, which track the synchronized trajectories of both features across the time-distance plots. It is particularly noteworthy that the slipping motion of the observed JBP provides a suggestive signature of slipping magnetic reconnection. Consequently, this may imply that the dynamical coupling of the photosphere and chromosphere mediated by slipping magnetic reconnection results from the persistent emergence and horizontal transport of magnetic flux within the light bridge.

\section{Conclution and Discussion}
Our multi-wavelength analysis of recurrent LB jets reveals a cross-layer dynamic process covering the solar atmosphere from the photosphere to the transition region. We find that the initiation and evolution of these LB jets are intrinsically linked to the photospheric magnetic dynamics of the sunspot LB. Our study provides clear evidence that some sunspot LB jets are driven by slipping magnetic reconnection. Morphologically, these LB jets share similar characteristics with other solar jets, exhibiting a collimated jet spire and a JBP.  In all six events studied here, the appearance of the JBP consistently precedes the LB jet ejection, a sequence analogous to that observed in coronal jets. The results show that all six JBPs exhibited pronounced slipping motions with a mean velocity of 1.1 $\speed$, accompanied by simultaneous corresponding slipping of the jet spires' leading edges in the upper chromosphere. Intriguingly, each slipping JBP was found to be temporally and spatially correlated with the appearance of photospheric horizontal motion, suggesting that these apparent motions likely trigger three-dimensional reconnection in the lower atmosphere. Spectroscopic signatures from CHASE further support the interpretation that magnetic reconnection triggers these LB jets. This observation not only provides strong evidence that some LB jets share a common formation mechanism with coronal jets, where both are driven by magnetic reconnection in the lower atmosphere, but also provides direct evidence of slipping reconnection occurring above sunspot LBs.
\par
A standard anemone jet often shows an inverted-Y shape consisting of a straight plasma beam and a bright point base, which is commonly interpreted as a result of magnetic reconnection between the emerging flux and the preexisting ambient coronal field \citep{1995Natur.375...42Y}. The inverted-Y morphology is not frequently visible due to the limited instrumental resolution. Small-scale jets are ejected from the JBP. Recently, similar base brightenings have been reported across a range of solar phenomena, from coronal jets \citep[e.g.,][]{2009SoPh..259...87N,2010ApJ...720..757M}, to microjets \citep[e.g.,][]{2021ApJ...918L..20H},  macrospicules  \citep[e.g.,][]{2023ApJ...942L..22D}, picoflare jets\citep{2023Sci...381..867C} and network jets \citep{2014Sci...346A.315T}. LB jets are accompanied by brightenings at their bases, sometimes even exhibiting the inverted-Y morphology mentioned in our study and the previous work by \citet{2018ApJ...854...92T}. In our work, the kinematic evolution of the LB jets is closely modulated by these JBPs' dynamics: the merging of two JBPs significantly enhances LB jet ejection, leading to increased extension lengths, while the fading of JBPs correlates with the weakening of the upward-ejecting LB jets. These results indicate that LB jets share a common physical mechanism with other solar jets ubiquitously observed across all solar regions, magnetic reconnection.

\par
Recent high-resolution observations and three-dimensional (3D) magnetic extrapolations have revealed that straight anemone jets are characterized by a fan–spine magnetic topology. Additionally, the round-trip slipping motion of the JBP of the fan-spine jet along the inner ribbons \citep{2012ApJ...760..101W} and circular flare ribbons \citep{2019ApJ...885L..11S} has been reported, providing evidence for the slipping magnetic reconnection process within quasi-separatrix layers (QSLs). QSLs, which are drastic changes in the magnetic field connectivity gradient but are still continuous, are considered ideal locations for slipping magnetic reconnection to occur. In solar eruptions, such slipping motion of footpoint brightenings is widely recognized as a consequence of the local diffusion in the magnetic reconnection region, where the neighboring magnetic field lines continually exchange connectivities \citep{2006SoPh..238..347A}. Given the magnetic shear between the horizontal fields of the LB and the vertical fields of the surrounding sunspot, the boundary of the LB serves as a favorable site for magnetic reconnection. 
\par
Previously, most studies have supported the view that the LB jets at fixed locations are driven by magnetic reconnection between emerging flux and the ambient sunspot field \citep[e.g.,][]{2017A&A...597A.127B, 2018ApJ...854...92T, 2020A&A...642A..44H}. However, the slipping motion of JBPs at the base of the LB jets, a critical dynamical feature, has rarely been reported \citep[e.g.,][]{2019ApJ...870...90B,2020ApJ...904...84L,2022A&A...667A..24L}, and observational evidence of slipping magnetic reconnection within the lower solar atmosphere remains remarkably sparse. This makes the current event a rare and critical case for understanding the dynamics of slipping magnetic reconnection in the lower atmosphere. 
In sporadic previous works, \citet{2019ApJ...870...90B} and \citet{2022A&A...667A..24L} reported potential signatures of slipping reconnection above LB, in which the photospheric convective motions play an important role in triggering the reconnection process. \citet{2020ApJ...904...84L} detected opposite magnetic polarity in a light bridge,  supporting the idea that the slipping reconstruction results from new flux emergence in the light bridge. However, the underlying driving mechanism of the slipping motion remained unaddressed in these studies. Building upon these findings, our work establishes the spatiotemporal relationship between the kinematic evolution of LB jets, the motion of slipping JBPs, and the photospheric horizontal motion. This comprehensive analysis provides evidence that the slipping motion of the JBP is driven by slipping magnetic reconnection, which is continually fueled by photospheric dynamics. In our study, the observed slipping motions of the JBPs serve as a compelling signature of 3D magnetic reconnection; notably, within the framework of 3D reconnection, these jet-associated brightenings are physically equivalent to slipping flaring kernels \citep[e.g.,][]{2013A&A...555A..77J,2015ApJ...804L...8L,2016ApJ...823...41D,2019ApJ...887..118C}.

\par
Based on the magnetic field configuration of the slipping reconnection in the lower atmosphere, we propose the following scenario, as illustrated in Figure 6 (A). In this schematic diagram, convective upflows in the northern LB transport horizontal magnetic fields to the surface layers, manifesting as the observed horizontal motion. The migration of these horizontal motions might be driven by the bent field lines connecting external vertical fields and internal horizontal fields, a process consistent with the simulations by \citet{2015ApJ...811..138T}. As these horizontal motion transport magnetic flux toward the LB boundaries, the carried horizontal fields gradually interact with the pre-existing vertical umbral fields. This interaction triggers slipping magnetic reconnection, which we suggest is responsible for the observed slipping motion of the JBPs. Subsequently, cool chromospheric material (the LB jets) is launched into the higher atmosphere by the tension force of the curved reconnected field, a process analogous to the sling-shot effect simulated by \citet{1995Natur.375...42Y}. Moreover, the enhanced emission at the leading edge of the jets in AIA 171 \AA~ may result from the compression between these jets and the overlying atmosphere \citep[e.g.,][]{2017ApJ...838....2Z, 2018ApJ...854...92T}.
\par 
The presence of convection upflows within the sunspot light bridge has been reported by previous simulations \citep[e.g.,][]{2015ApJ...811..137T,2015ApJ...811..138T} and observations \citep[e.g.,][]{2010ApJ...718L..78R,2018ApJ...865...29Z}. The diverging convective outflow carries the magnetic flux to the LB, potentially turning into horizontal motion. In our study, we measured horizontal motion speeds ranging from 1.3-6.5 $\speed$. These values are comparable to the horizontal unidirectional motion speed of $\sim$8 $\speed$ along the boundary of the light bridge, as simulated by \citet{2015ApJ...811..138T}, and are also consistent with the maximum speed of 1.8 $\speed$ in observation reported by \citet{2019ApJ...870...90B}. 
\par
As proposed by \citet{2018ApJ...854...92T}, jets above sunspot LBs comprise two distinct components: persistent, gentle short jets are related to the upward leakage of magnetoacoustic waves from the photosphere, and occasionally, longer, and faster jets driven by intermittent magnetic reconnection. These occasionally occurring reconnection-driven jets are often superimposed on these gentle, persistent jets, creating an intermittent but more intense ejected component of the jets. The jets driven by magnetic reconnection generally exhibit high velocities ($\sim$40-100$\speed$) \citep[e.g.,][]{2016A&A...590A..57R,2020ApJ...904...84L}. But in our observation, the jets exhibit speeds of $\sim$16.3 $\speed$, which is smaller than the previous work. On one hand, the relatively low observed velocities may be attributed to a combination of projection effects and the 45s temporal resolution of the NVST. On the other hand, when evaluating the driving mechanisms, contributions from convective-generated slow-mode magnetoacoustic waves \citep[e.g.,][]{2018ApJ...854...92T} and leaked p-mode waves from the photosphere \citep[e.g.,][]{2015ApJ...804L..27Y, 2017ApJ...848L...9H} cannot be entirely excluded. In our analysis, the propagation directions of slow-mode magnetoacoustic waves and leaked p-mode waves differ from the slipping direction of the JBPs, appearing nearly perpendicular to each other. As a result, whether these waves exert a negative influence on the slipping magnetic reconnection process, and thus contribute to the observed slow LB jets velocity, remains an open question.

\begin{acknowledgments}
The authors would like to express their sincere gratitude to Prof. Hui Tian and his team (PKU) and Assoc. Prof. Hechao Chen (YNU) for their insightful and stimulating discussions. This work is supported by the Strategic Priority Research Program of the Chinese Academy of Sciences,  Grant No. XDB0560000, NSFC grant 12403065. X.L.Y was supported by the National Science Foundation of China (NSFC) under Nos. 12325303, the Yunling Scholar Project (Yunnan Revitalization Talent Support Program), Yunnan Key Laboratory of Solar Physics and Space Science under No. 202205AG070009, and Yunnan Fundamental Research Projects under Nos. 202301AT070347 and 202301AT070349. J.C.W. was supported by the Yunnan Fundamental Research Projects under Nos. 202501AW070002. Z.X. was supported by Yunnan Fundamental Research Projects under Nos.202301AT070349. Y.Y.X. was supported by Yunnan Fundamental Research Projects under Nos.202301AT070350. As one of the primary observing facilities of the Fuxian Solar Observatory, NVST is jointly operated and administered by Yunnan Observatories and the Center for Astronomical Mega-Science, Chinese Academy of Science. This work also uses data from the CHASE mission supported by the China National Space Administration. 
\end{acknowledgments}

\begin{figure}[ht!]
\centering
\includegraphics[width=0.9\textwidth]{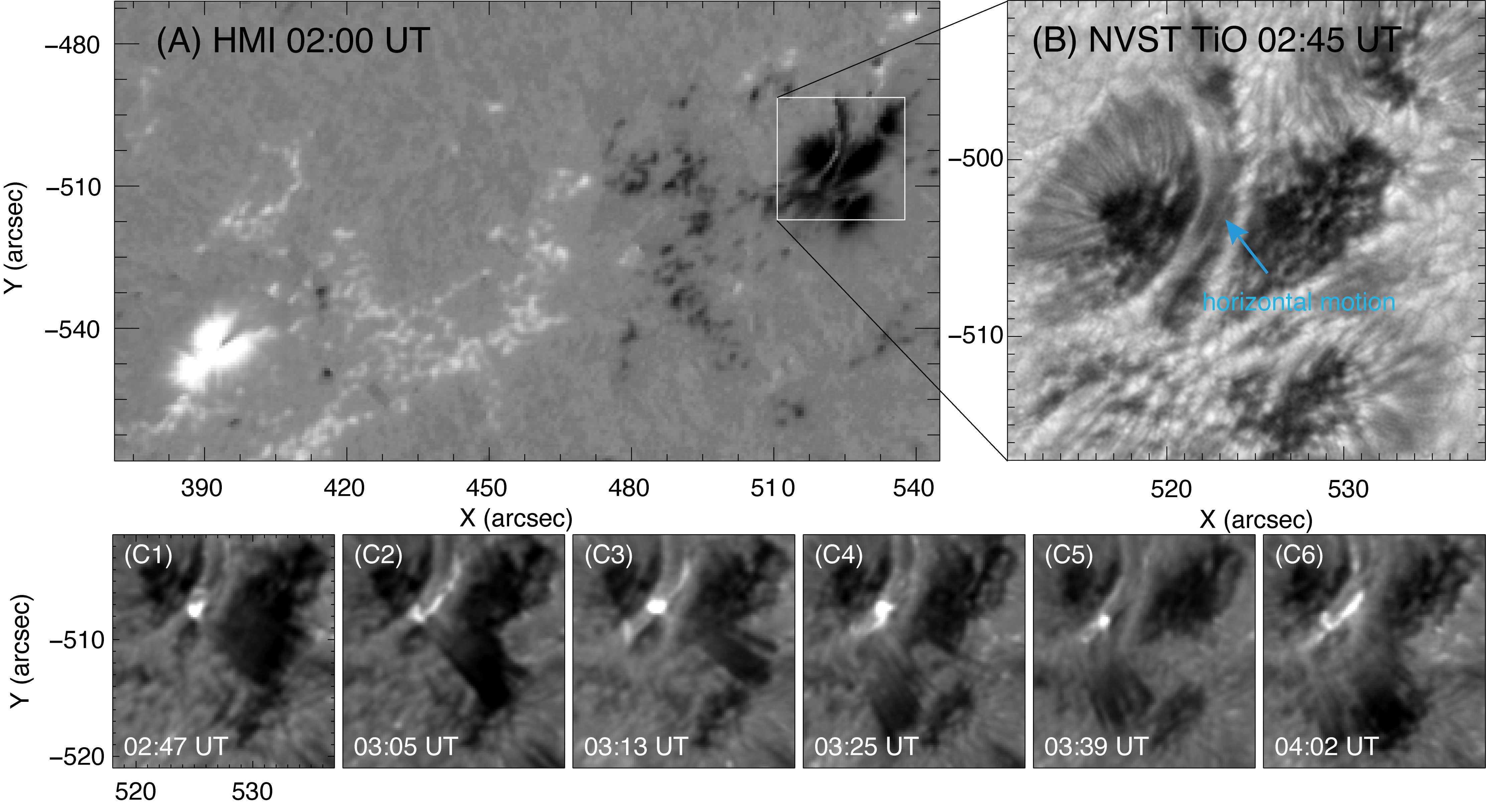}
\caption{Overview of the sunspot, light bridge, and LB jets in AR 14153 on 2025 July 30. (A) The longitudinal magnetic field at 02:00 UT shows the magnetic environment of the source region. (B) NVST TiO image at 02:45 UT, and the field of view (FOV) is indicated by the white box in (A). The cyan arrow denotes the horizontal motion along the light bridge. (C1)$-$(C6) NVST H$\alpha$ -0.6\AA~images show six of the LB jets ejected along the light bridge. An animation of the whole dynamic process of the repeated LB jets is available from 02:41 UT to 04:18 UT. The duration of the animation is 6s.
\label{fig:general}}
\end{figure}

\begin{figure}[ht!]
\plotone{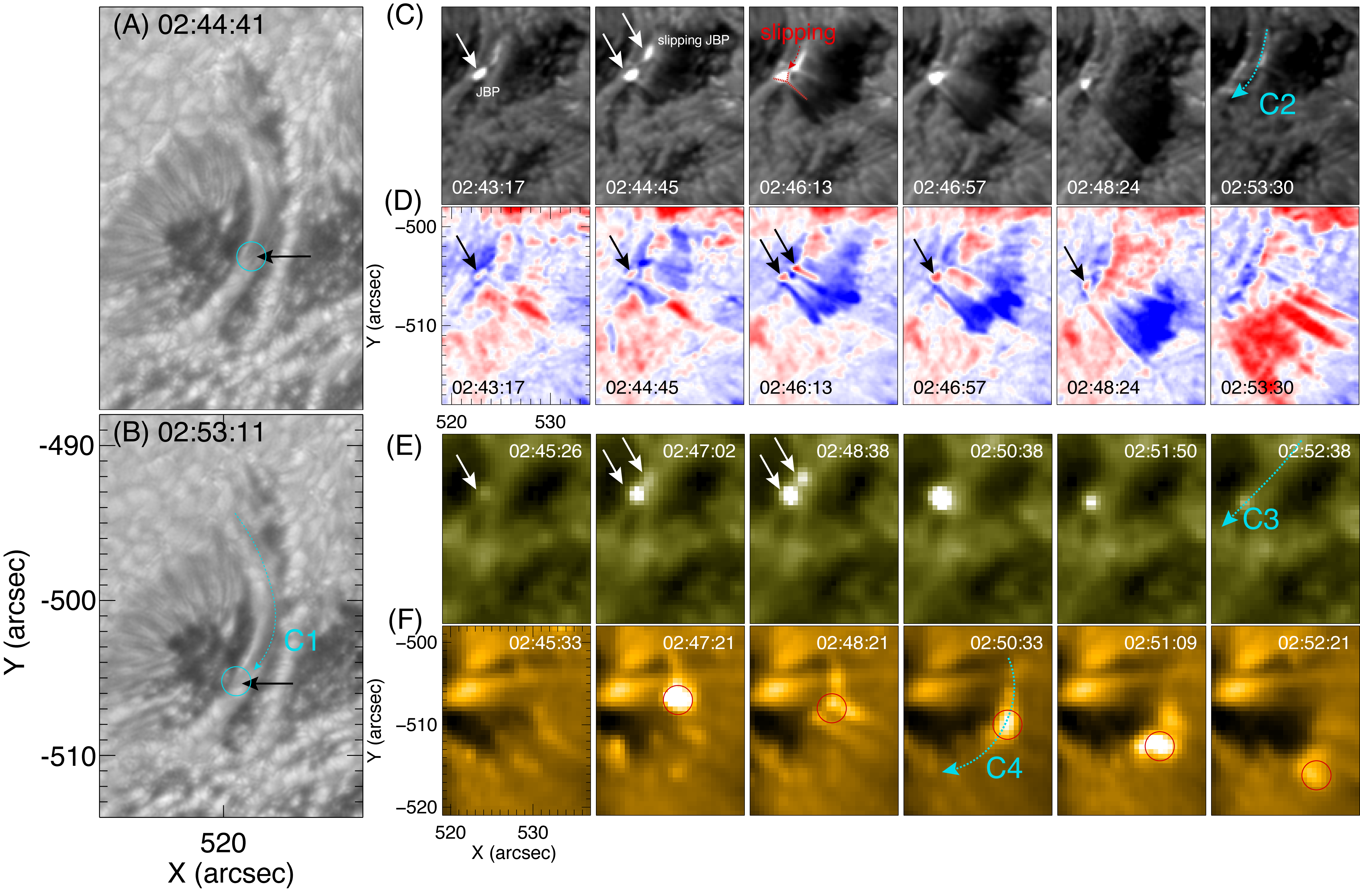}
\caption{(A)$-$(F) Zoomed-in snapshots showing the evolution of an LB jet from 02:41 UT to 02:50 UT in NVST TiO, H$\alpha$ -0.6 \AA~,  AIA 1600 \AA~, 171\AA~images and Doppler proxy maps. The black arrows and the cyan circles in panels (A) and (B) denote a horizontal motion front along the light bridge. The white and black arrows in panel (C)$-$(E) point out the bright footpoints at the base of the LB jet. The red circle marks the bright plasma moving along the leading edge of the LB jet. The dashed cyan arrows in panels (B), (C), (E), and (F) denote the slit position used for the time–distance plot shown in Figure 5 (A1)$-$(A4). NVST H$\alpha$ line wing images (panels C) and the AIA 1600 \AA~and 171 \AA~(panels E and F) are shown in the animation.
\label{fig:general}}
\end{figure}

\begin{figure}[ht!]
\plotone{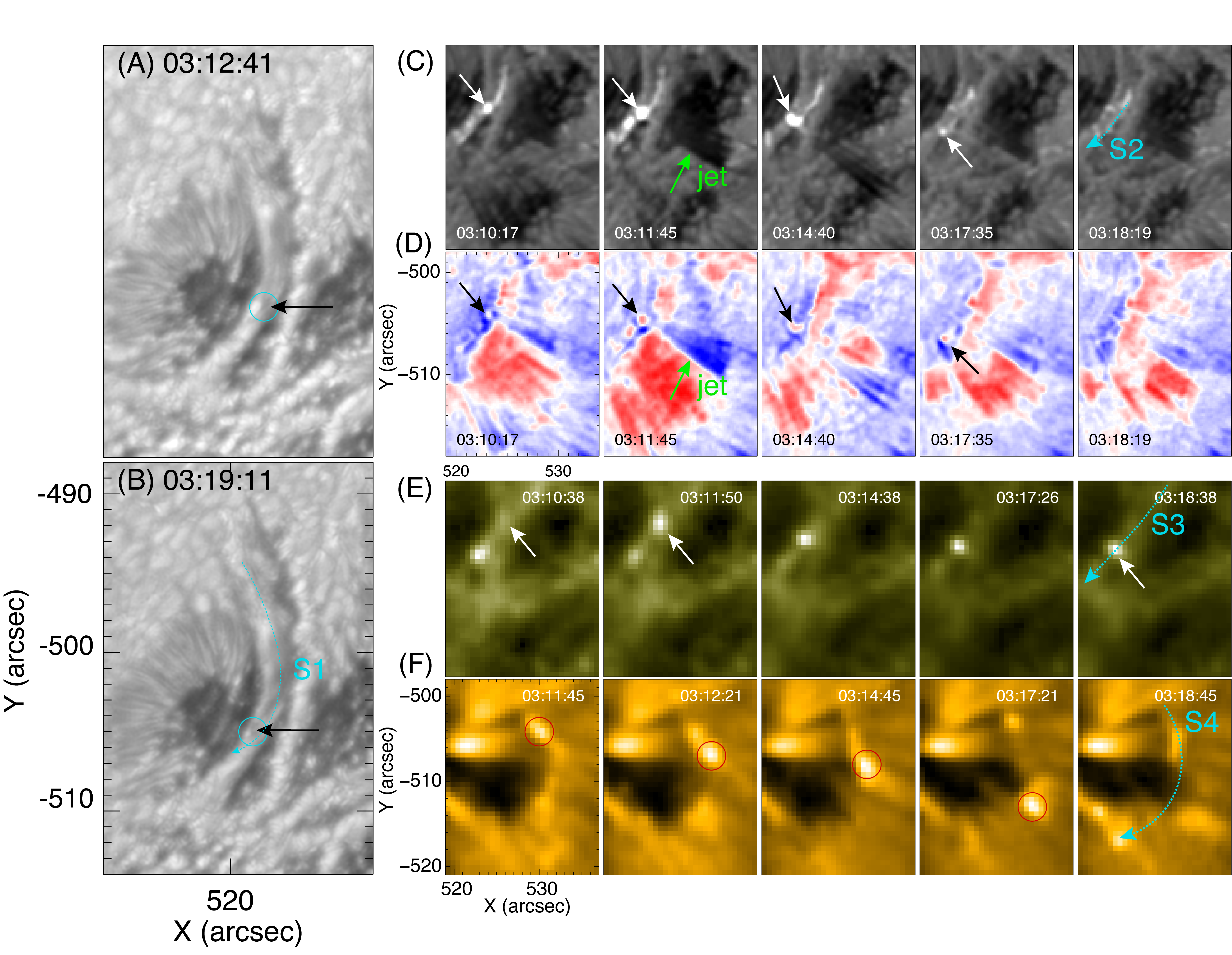}
\caption{Similar to Figure 2, but from 03:08 UT to 03:19 UT in panels (A)$-$(F). The black arrows and the cyan circles in panels (A) and (B) denote the horizontal motion front. The white and black arrows in panel (C)$-$(E) point out the bright footpoints. The red circle marks the bright plasma. The slit position for the time-distance plot in Figure 5 (B1)$-$(B4) is shown by the dashed cyan arrows in panels (B), (C), (E), and (F). NVST H$\alpha$ line wing images (panels C) and the AIA 1600 \AA~and 171 \AA~(panels E and F) are shown in the animation.
\label{fig:general}}
\end{figure}

\begin{figure}[ht!]
\centering
\includegraphics[width=0.7\textwidth]{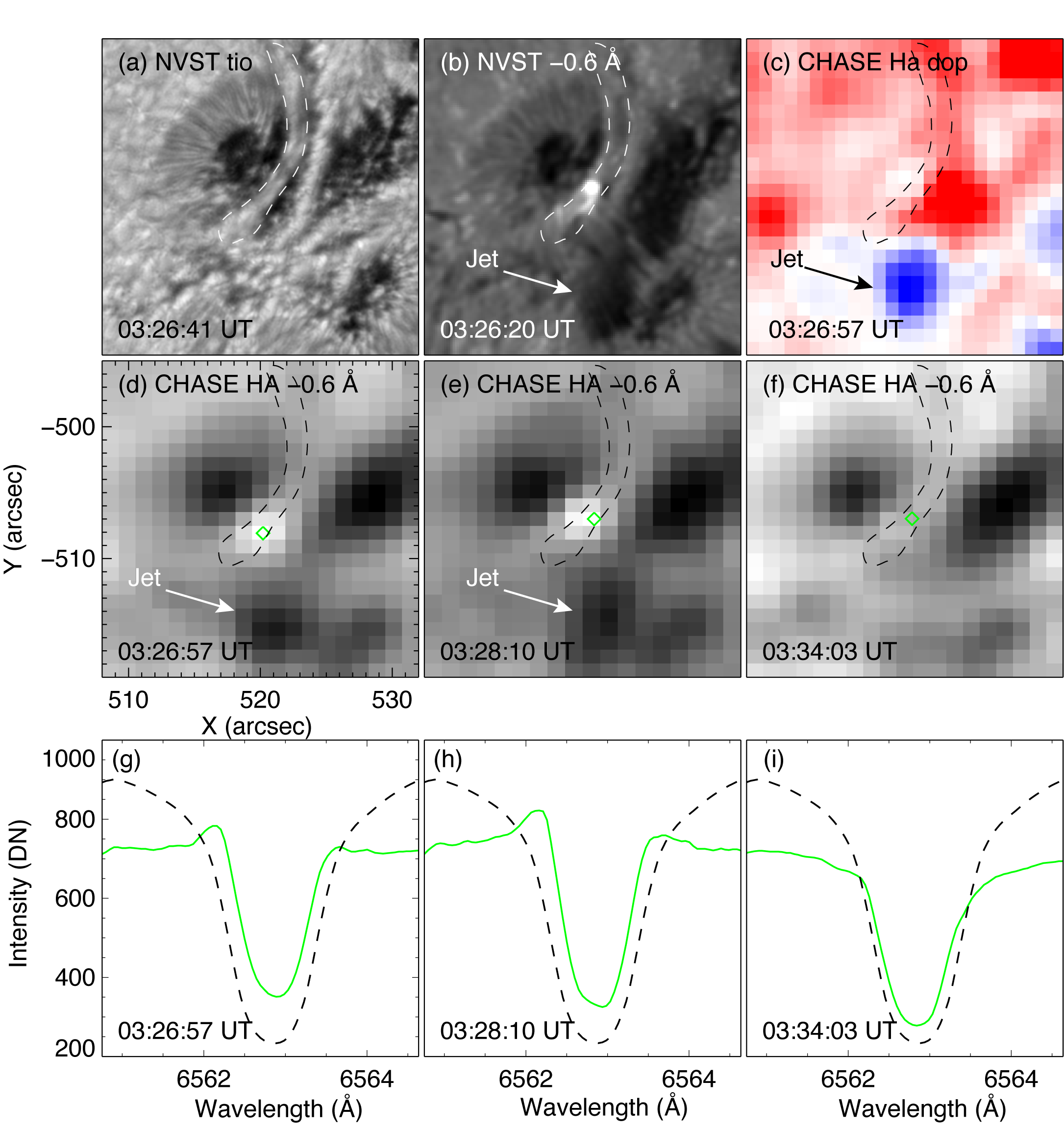}
\caption{The dashed curves in the CHASE Dopper map, NVST, and CHASE images show the light bridge of our interest. The green H$\alpha$ profiles of selected points in panels (g)$-$(i) are marked with green diamond symbols in (d)$-$(f). And the dashed lines represent the reference H$\alpha$ profile. 
\label{fig:general}}
\end{figure}

\begin{figure}[ht!]
\centering
\includegraphics[width=0.9\textwidth]{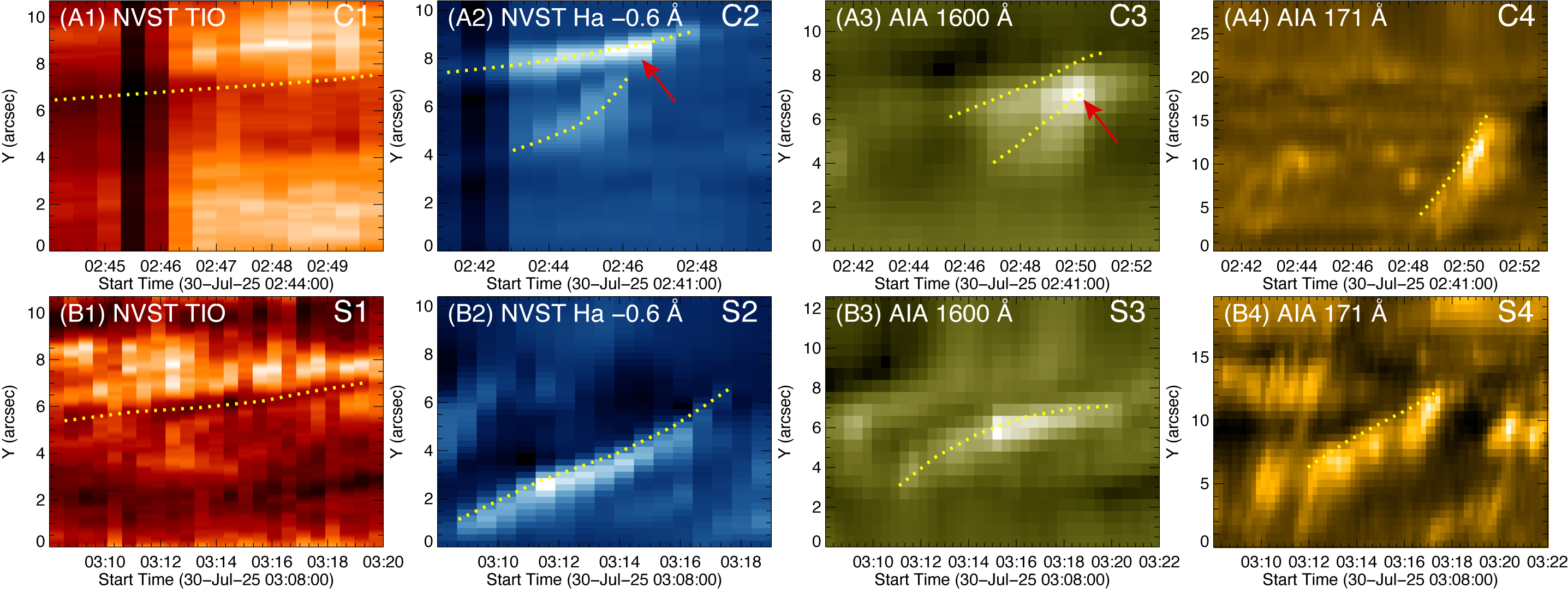}
\caption{Time-distance map along slices C1$-$C4 and S1$-$S4 shown in Figures 2 and 3. Panels (A1) and (B1) show the horizontal motion in the LB at different times. Panels (A2), (B2), (A3), and (B3) show the slipping bright footpoints in the LB for the different images of NVST H$\alpha$ off-band and AIA 1600 \AA~. The red arrows in panels (A2) and (A3) mark the merged position between a bright patch and a slipping bright footpoint. Panels (A4) and (B4) display the moving plasma at the leading edge of the LB jets in the AIA 171 \AA~channel. 
\label{fig:general}}
\end{figure}

\begin{deluxetable*}{lccccccccccc}
\tabletypesize{\scriptsize}
\setlength{\tabcolsep}{3pt} 
\tablewidth{0pt} 
\tablecaption{Basic properties of the LB jet Events \label{tab:deluxesplit}}
\tablehead{
 \colhead{\splitcell{Event\\ID}} & \colhead{\splitcell{Telescope/\\Passbands}} & \colhead{\splitcell{Time\tablenotemark{a}\\(UT)}} & \colhead{\splitcell{Corresponding \\ CHASE\\Time\\(UT)}} & \colhead{\splitcell{BP\tablenotemark{a}\\ Slipping\\Duration\\(min)}} &
  \colhead{\splitcell{Motion\tablenotemark{b}\\Velocity\\($\speed$)}} &\colhead{\splitcell{BP\tablenotemark{c}\\Slipping\\Velocity\\($\speed$)}} &\colhead{\splitcell{LB jet\tablenotemark{d}\\Slipping\\velocity\\($\speed$)}} & \colhead{\splitcell{LB jet\tablenotemark{e}\\Ejected\\velocity\\($\speed$)}} &  \colhead{\splitcell{Maximum\tablenotemark{e} \\ Length/\\ (Mm)}}\
} 
\startdata 
C1 & NVST H$\alpha$ &  02:41$-$02:50&  \dots& 9& 2.3& 1.5  &  3.7  &  24 &  9.3 \\ 
\hline
C2 & NVST H$\alpha$ &   02:50$-$03:08&     \dots&7& 4.3&0.8 &1.9    &  18 & 9   \\ 
\hline
C3& NVST H$\alpha$ &    03:08$-$03:19&     \dots&10&6.5& 1.4 &  1.1  & 10  &  8.6  \\ 
\hline
C4& NVST H$\alpha$ &    03:19$-$03:32&   03:26$-$03:32 & 13& 3.6& 1 &  0.8  &  12 & 7.1  \\ 
\hline
C5& NVST H$\alpha$ &    03:32$-$03:42&   03:32$-$03:42&8& 3.3&1& 0.9&  14  &   6.2   \\ 
\hline
C6& NVST H$\alpha$ &    03:42$-$04:18&    03:42$-$03:55& 26&1.3& 0.6 &  1.1  & 19.5  & 8.7   \\ 
\hline
\enddata
\tablecomments{The date of observation is from 2025 July 30.}
\tablenotetext{a}{The Time represents the time period during which the LB jets occur in the H$\alpha$ blue wing of NVST. }
\tablenotetext{b}{The speeds of horizontal motion derived from a fitting linear function in NVST TiO image as displayed in Figure 6 (e).}
\tablenotetext{c}{The slipping speeds of bright points come from a fitting linear function in NVST H$\alpha$ -0.6\AA~.}
\tablenotetext{d}{The slipping speeds of the plasma at the leading edge of LB jets obtained from a fitting linear function in AIA 171 \AA~, see the example Case 1 and Case 3 in Figures 5 (A4) and (B4).}
\tablenotetext{e}{In NVST H$\alpha$ -0.6\AA~, the maximum length reached during each ejection along the LB jet determines the ejection speed.}
\end{deluxetable*}

\begin{figure}[ht!]
\centering
\includegraphics[width=0.9\textwidth]{fig6.png}
\caption{(a) NVST TiO image at 03:26 UT. (b) The SDO/HMI vector magnetogram at 03:48 UT. Green arrows represent the strength and direction of the transverse magnetic fields. The boundaries between the sunspot and the LB are highlighted by red circles. (c) NVST H$\alpha$ off-band at 03:01 UT. (d) The field of view in panel (d) corresponds to the cyan box indicated in panel (c), showing the photospheric flow fields derived via the optical flow method from NVST TiO images. (e)$-$(f) Time-distance diagrams along the slices AB and CD shown in panels (a) and (b). The yellow arrows point to the six horizontal motions and the JBPs. (A) Schematic depiction of the slipping processes of the LB jet originating from an LB.  Cyan arrows illustrate the diverging convective flows toward the LB. Green lines denote the ambient magnetic field lines anchored in the surrounding umbral region. The dark line signifies a horizontal magnetic field line being transported by the convective flow.  Red star symbols mark the sites of slipping magnetic reconnection between the transported horizontal flux and the umbral field, which triggers the LB jets (pink shaded area); the slipping motion of the jet is indicated by the blue arrow.
\label{fig:general}}
\end{figure}

\bibliography{reference}{}
\bibliographystyle{aasjournalv7}

\end{document}